\documentclass[aps,prb,10pt, twocolumn, superscriptaddress, longbibliography, nobalancelastpage]{revtex4-1}
\usepackage{graphicx}
\usepackage{amsmath}
\usepackage{amssymb}
\usepackage{amsfonts}
\usepackage{dsfont}
\usepackage{braket}
\usepackage{color}
\usepackage{braket,slashed}
\usepackage[mathscr]{euscript}
\definecolor{darkblue}{rgb}{0, 0, 0.8}
\usepackage[colorlinks=true, breaklinks=true, linkcolor=red, citecolor=blue, urlcolor=blue]{hyperref} 
\usepackage{hyperref}
\usepackage{subfigure}
\usepackage{xfrac}
\usepackage{bm}
\usepackage{kantlipsum}
\usepackage{enumitem}
\usepackage{tikz}
\usepackage{framed}
\usepackage{graphicx}
\usepackage{subfigure}

\allowdisplaybreaks[1]

\newcommand{\code}[1]{\texttt{#1}}

\DeclareMathOperator*{\argmin}{arg\ min}

\newcommand{\bA}{\ensuremath{\bar{A}}}

\begin{document}

\title{A scaling hypothesis for projected entangled-pair states}

\author{Bram Vanhecke}
\email{bavhecke.Vanhecke@UGent.be}
\affiliation{Department of Physics and Astronomy, University of Ghent, Krijgslaan 281, 9000 Gent, Belgium}

\author{Juraj Hasik}
\email{j.hasik@uva.nl}
\affiliation{Institute for Theoretical Physics, University of Amsterdam, Science Park 904, 1098 XH Amsterdam, The Netherlands}

\author{Frank Verstraete}
\affiliation{Department of Physics and Astronomy, University of Ghent, Krijgslaan 281, 9000 Gent, Belgium}

\author{Laurens Vanderstraeten}
\affiliation{Department of Physics and Astronomy, University of Ghent, Krijgslaan 281, 9000 Gent, Belgium}

\begin{abstract}
	We introduce a new paradigm for scaling simulations with projected entangled-pair states (PEPS) for critical strongly-correlated systems, allowing for reliable extrapolations of PEPS data with relatively small bond dimensions $D$. The key ingredient consists of using the effective correlation length $\xi$ for inducing a collapse of data points, $f(D,\chi)=f(\xi(D,\chi))$, for arbitrary values of $D$ and the environment bond dimension $\chi$. As such we circumvent the need for extrapolations in $\chi$ and can use many distinct data points for a fixed value of $D$. Here, we need that the PEPS has been optimized using a fixed-$\chi$ gradient method, which can be achieved using a novel tensor-network algorithm for finding fixed points of 2-D transfer matrices, or by using the formalism of backwards differentiation. We test our hypothesis on the critical 3-D dimer model, the 3-D classical Ising model, and the 2-D quantum Heisenberg model.
\end{abstract}

\maketitle

Although variational algorithms over the class of projected entangled-pair states (PEPS) \cite{Verstraete2004, Cirac2020} are by now well established for simulating  ground states of 2-D quantum spin systems \cite{Jordan2008, Vanderstraeten2016, Corboz2016} and free energies of 3-D statistical mechanical systems \cite{Nishino2001, Vanderstraeten2018}, the accurate simulation of critical points and gapless phases in (2+1)-D remains elusive due to the prohibitive computational cost of simulating PEPS with a large bond dimension. Indeed, PEPS are the natural higher-dimensional generalization of matrix product states (MPS), but, unlike for MPS, expectation values cannot be computed exactly as that would require the contraction of an infinite two-dimensional tensor network. Instead, such networks are typically contracted by constructing effective environments -- either as a boundary MPS \cite{Verstraete2004, Fishman2018} or with corner-transfer matrices \cite{Nishino1996, Orus2009, Corboz2011} -- both of which carry a bond dimension $\chi$ controlling the error in the environment, and provide an effective environment of similar accuracy for the same $\chi$.

\par The investigation of critical systems by PEPS has been severely hampered by the presence of these two control parameters. 
In the case of MPS, it has been realized that working at finite bond dimension induces a relevant perturbation in the system, much in the same way as finite sizes enter in exact diagonalization or Monte-Carlo techniques. As an analogue to finite-size scaling \cite{Brezin1988, Cardy1984, Fisher1972}, the theory of finite-entanglement scaling \cite{Nishino1996b, Pollmann2009, Tagliacozzo2008, Pirvu2012, Vanhecke2019} identifies an induced length scale due to the entanglement compression caused by finite $\chi$. This has allowed MPS simulations to reach a high precision in the determination of critical points and exponents, both for 1-D quantum systems and problems in 2-D statistical mechanics.

\par Recently, there have been two crucial developments in realizing the same program for PEPS simulations. First, variational algorithms were devised that allow to find an optimal PEPS approximation for a given model Hamiltonian \cite{Vanderstraeten2016, Corboz2016} or transfer matrix \cite{Nishino2001, Vanderstraeten2018, Vanhecke2020} at a certain value of the bond dimension $D$ -- in contrast to the simple-update or full-update schemes that were shown \cite{Vanderstraeten2016} to yield suboptimal PEPS results. Second, the correlation length of the optimal PEPS at a certain bond dimension, $D$, was identified as a suitable length scale for extrapolating variational energies and other observables \cite{Rader2018, Corboz2018, Czarnik2019}. In both these advances, the role of the environment bond dimension $\chi$ was not given any meaningful content: one simply assumes that all optimizations are done for large enough values of $\chi$, and correlation lengths are then extrapolated to get rid of any $\chi$ dependence\cite{rams2018}. This makes PEPS simulations very costly, as the $\chi$-limit is often very tough if not impossible to reach, especially for large $D$. Moreover, as an implication, only a very limited number of data points can contribute to the extrapolation -- i.e., a single point for every value of $D$ accessible by computational resources. Contrary to MPS, which can be optimized up to bond dimensions of order $O(10^5)$, the gradient-based PEPS optimizations have so far been limited to $D<10$.

\par In this paper, we place finite-$\chi$ and finite-$D$ data on an equal footing in PEPS extrapolation schemes. In order for this to be possible, the optimized PEPS must be an extreme point of the cost function for a given value of $D$ and $\chi$. That is, we need to solve the optimization problem
\begin{equation} \label{eq1}
	A = \argmin_A f_{\{D,\chi\}}(A,\bar{A})
\end{equation}
for tensor(s) $A$ forming the PEPS ansatz where the cost function is the value of the energy (quantum) or free energy (classical) as a result from the approximate PEPS contraction \emph{at a fixed value of $\chi$}. In the case of two-dimensional PEPO transfer matrices, this can be achieved by gradient optimization, since the $\chi$-dependent gradient of the cost function is obtained by a simple contraction (a summary of the requisite methods is presented in the supplementary material for a brief description, which includes Ref.~\onlinecite{Vanderstraeten2019}). For quantum Hamiltonians, one can construct a complex time evolution PEPO\cite{Vanhecke2020} and do the same, or the optimization can be done by supplementing existing contraction algorithms with a backwards differentiation step, as e.g. done using automatic differentiation \cite{Liao2019, Hasik2020}.

\par The crucial contribution of this paper is the following hypothesis: close to a critical point the dependence of expectation values $\braket{O}$ of an optimized PEPS wavefunction -- optimized in the sense of Eq.~\ref{eq1} -- as a function of $D$ and $\chi$ exhibits a data collapse when expressed through the induced correlation length $\xi$ of the environment\footnote{For an exact environment this $\xi$ would also be the correlation length of the PEPS.} --the one used during optimization-- with bond dimension $\chi$:
\begin{equation}
	\braket{O}(D,\chi) = \braket{O}(\xi(D,\chi)).
\end{equation}
The consequences of the above hypothesis are twofold. On the one hand, it implies that even optimized PEPS with relatively small $\chi$ can be used for accurate extrapolations, provided that they are in the scaling regime. On the other hand, it allows for a substantial increase in the amount of data relevant for extrapolation as much more data points can be constructed by varying $\chi$ for each $D$ considered. We want to stress that this hypothesis only holds for PEPS algorithms that yield extremal points for the cost function in Eq.~\ref{eq1}, a condition that is not satisfied for simple and full-update algorithms and for earlier variational algorithms \cite{Vanderstraeten2016, Corboz2016}, but which is satisfied for the exact $(D,\chi)$-gradient algorithms such as the one presented in the supplemental and the ones based on backwards differentiation.

\begin{figure}
	\includegraphics[width=\linewidth]{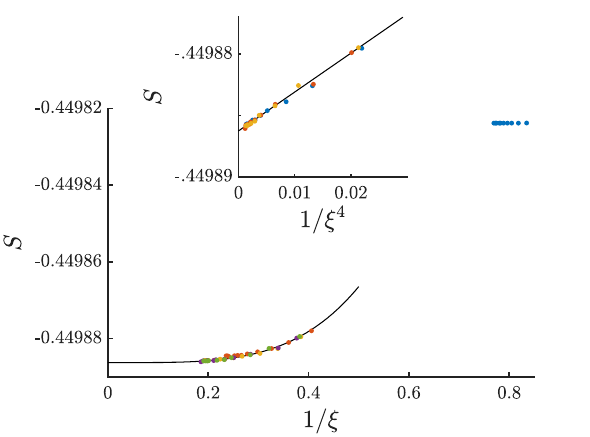}
	\caption{Residual entropy per site for the dimer covering problem on the cubic lattice. We have optimized PEPS tensors for $D=(2,3,4,5,6)$ and $\chi=(20, 26, 34, 44, 58, 75, 97, 126, 164, 213, 276)$; the same MPS bond dimension was used for the contraction of the double- and triple-layer (see supplementary material). The correlation length is extracted from the double-layer boundary MPS.  A fit (black) on the $D>3$ data reveals a clear $x^4$ power law (see inset) for which the origin is to us an enigma.}
	\label{fig:dimers}
\end{figure}

\begin{figure}
	\includegraphics[width=\linewidth]{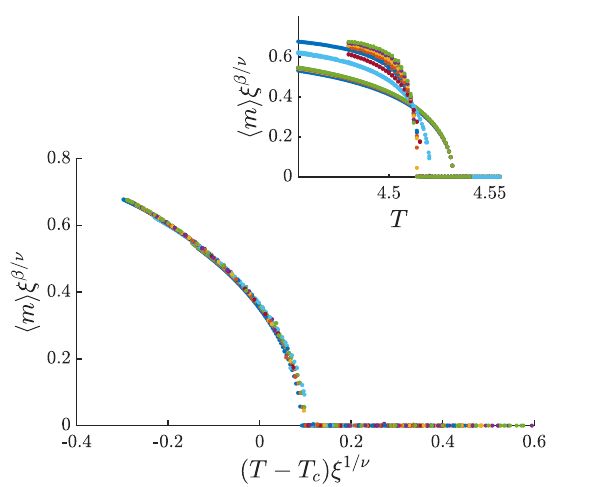}
	\caption{Magnetization of the classical 3-D Ising model. We have optimized PEPS tensors for $D=(2,3,4)$ and $\chi=(10,15,20,25,30,40,50)$ for 100 equally spaced temperatures around the critical temperature. The same MPS bond dimension was used for the contraction of the double- and triple-layer (see supplementary material). To plot the data, we shift the temperature by $T_c=4.511528$ (obtained from Monte-Carlo simulations \cite{Talapov1996}) and rescaled both axes with appropriate powers $\beta=0.326419$ and $\nu=0.629971$ of $\xi$ (obtained from the conformal bootstrap method \cite{Simmons-Duffin2015}), with $\xi$ the correlation length of the double-layer MPS environment. In the inset we provide a partial data collapse (also showing the abundance of data points), showing the crossing of the rescaled magnetization at the critical temperature.}
	\label{fig:3dising}
\end{figure}

\par We first illustrate this hypothesis by scaling the free energy and order parameter of statistical mechanical problems in three dimensions, where we optimize PEPS fixed points of the 2-D transfer matrix \cite{Vanderstraeten2018} with a gradient optimization at fixed values of $D$ and $\chi$. We choose the $\chi$ values of the double and triple layer equal, but we expect that this is not a strict requirement. In fact it might even be a good idea to choose them differently), hence expanding the set of data points even further. First, we consider the partition function of dimer coverings on a cubic lattice. This problem was first tackled with PEPS in Ref.~\onlinecite{Vanderstraeten2018}, but here we optimize at fixed values of $D$ and $\chi$ allowing for better convergence of the PEPS optimization in addition to enabling us to test our scaling scheme. The data is condensed on a single plot in Fig.~\ref{fig:dimers}. We observe that, indeed, the PEPS data for different $(D,\chi)$ all lie on the same curve, even for relatively small values of $\chi$. An extrapolation in $1/\xi^4$ (motivated by the inset) gives rise to an extremely precise value for the residual entropy, $S=0.449886267(25)$, and has to be compared with the Monte-Carlo result \cite{Beichl1999} of $S=0.4466\pm0.0006$ -- the accurate determination of absolute quantities with Monte Carlo are well known to be very challenging. Next, we study the 3-D classical Ising model near criticality, for which we perform a collapse of all our $(D,\chi)$ data for the magnetization, shown in Fig.~\ref{fig:3dising}. To perform the data collapse, we use the critical exponents calculated by bootstrap \cite{Simmons-Duffin2015} and find the temperature $T_c$ that optimizes the collapse --in a total least squares sense--: $T_c=4.51170$; this should be compared to $T_c=4.511528(6)$ from Monte-Carlo simulations \cite{Talapov1996}\footnote{This particular way of exploiting the scaling hypothesis seems to be poorly conditioned to determine the precise values of the critical exponents, as opposed to the determination of $T_c$, which seems to be fairly robust. If we also fit the exponents, we find $T_c=4.5104$, and $\beta=0.3112$, $\nu=0.7375$. There are, of course, better ways to estimate the critical exponents. Note that while a value for $T_c$ can be straightforwardly extracted from a collapse, there is no canonical way to estimate the error bar.}. In these simulations we initialized PEPS' with other converged results and added a small perturbation to prevent getting stuck in a local minimum.

% \begin{figure}
	% \subfigure{\includegraphics[width=\linewidth]{./Figures/2dising1.pdf}} \\
	% \subfigure{\includegraphics[width=0.49\linewidth]{./Figures/2dising2.pdf}} 
	% \subfigure{\includegraphics[width=0.49\linewidth]{./Figures/2dising3.pdf}} 
	% \caption{Data for the 2-D transverse-field Ising model at the approximate \cite{Blote2002} critical field $h = 3.04438$, for $D=(2,3,4)$ and $\chi=(8,16,24,48,96,192)$. Top: energy vs. inverse correlation length of the environment MPS. Bottom-left: energy vs. the cube of the inverse correlation length. Bottom-right: magnetization vs that same inverse correlation length, both raised to appropriate powers to make quantities of unit scaling dimension. Fits are performed on the $D>2$ data and are shown in black.}
	% \label{fig:qising}
	% \end{figure}

\begin{figure}
	\subfigure{\includegraphics[width=\linewidth]{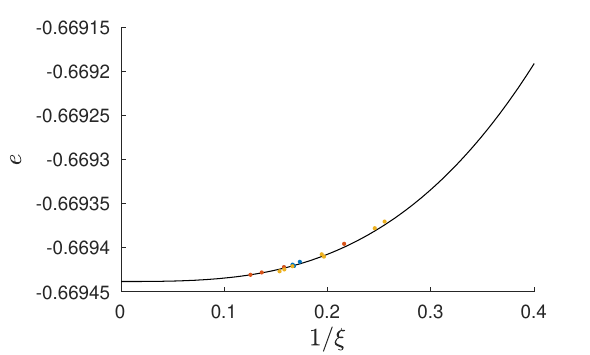}} \\
	\subfigure{\includegraphics[width=0.49\linewidth]{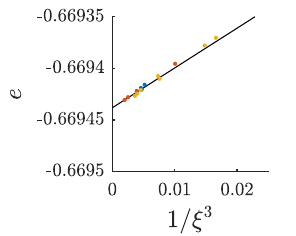}}
	\subfigure{\includegraphics[width=0.49\linewidth]{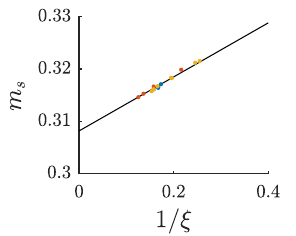}}
	\caption{Data for the 2-D Heisenberg model for $D=(6,7,8)$ and selected $\chi$ from 17 up to 200 for $D=6,7$ and 147 for $D=8$. Top: energy vs. inverse correlation length of the environment MPS. Bottom-left: energy vs. the cube of the inverse correlation length. Bottom-right: magnetization vs that same inverse correlation length. Fits are performed on all the data and are shown in black.}
	\label{fig:heisenberg}
\end{figure}

% \par Next we demonstrate the scaling hypothesis on two 2-D quantum systems, the transverse-field Ising model and the spin-$1/2$ antiferromagnetic Heisenberg model. In both cases, the PEPS is completely described by a single real tensor $A$ with $C_{4v}$ point group symmetry \cite{Hasik2020}, so we can use the original straightforward and highly-efficient formulation of the corner-transfer matrix algorithm \cite{Nishino1996}. In order to evaluate the gradient of the variational energy, the whole process of evaluating the energy expectation value is differentiated by backwards differentiation \cite{Liao2019, Hasik2020, Hasik2021}.
% \par We perform simulations for transverse-field Ising model at the approximate \cite{Blote2002} critical value of the field, see Fig.~\ref{fig:qising}. We test that this is indeed the critical point by extrapolating the magnetization, for which we find $m=0.00035(15)$. We perform this extrapolation, shown in  making explicit use of the same $\beta$ and $\nu$ used in the 3-D Ising case. We also extrapolate the ground state energy using a $\xi^{-3}$ law \cite{Rader2018}, to the value $e=-3.2342626(16)$. This compares nicely with the value $e=−3.2342623$, obtained by an infinite-$\chi$ PEPS extrapolation \cite{Rader2018}.

\par Finally we consider the square-lattice quantum Heisenberg model. Here, the PEPS wavefunction for the ground state is completely described by a single real tensor $A$ with $C_{4v}$ point group symmetry \cite{Hasik2020}, so we can use the original straightforward and highly-efficient formulation of the corner-transfer matrix algorithm \cite{Nishino1996}. In order to evaluate the gradient of the variational energy, the whole process of evaluating the energy expectation value is differentiated by backwards differentiation \cite{Liao2019, Hasik2020, Hasik2021}. Again, we optimize PEPS tensors for different $(D,\chi)$, and plot the energy and magnetization as a function of inverse correlation length in Fig.~\ref{fig:heisenberg}. The staggered magnetization can be extrapolated as a linear function of $1/\xi$, as was noted in Ref.~\onlinecite{Rader2018}, and is here motivated by a plot in Fig.~\ref{fig:heisenberg}; we find the extrapolated value $m_s=0.30771(31)$. The energy can be extrapolated in $1/\xi^3$, as shown in the left-bottom plot of Fig.~\ref{fig:heisenberg}, to $e=-0.6694401(10)$. These values should respectively be compared to the Monte-Carlo \cite{Sandvik1997,Sandvik2010} values $e=-0.669437(5)$ and $m_s=0.30743(1)$. %$e=−0.669437(5)$ and $m_s=0.30743(1)$. 
For the Heisenberg model, we impose $U(1)$ symmetry on the PEPS (not on the environment) and select the class that previously proved to be most effective \cite{Hasik2020}.

\par Notice that the $D=2$ data (blue indicators) for the dimer model are clearly not in the scaling regime. We have included them as an illustration of the fact that high enough bond dimensions must be chosen to reach the scaling regime. For all the above results we implement some filtering of the data, imposing that the magnetization decreases monotonically in $D$ and $\chi$ separately\footnote{Note that such a filtering on the cost function (residual entropy for the dimers, free energy for the 3D Ising model, and energy for the Heisenberg model) is not allowed, as it is not guaranteed to be monotonic in $\chi$ or $D$, and there are indeed data points with a lower value of the cost function than the extrapolated value.}. This allows us to identify data points that seemed completely converged, based on the norm of the gradient, but are in fact stuck at a fake minimum and also do not satisfy the scaling hypothesis. By restarting VUMPS with a random MPS it became clear that the MPS environment of this data was actually in a local maximum of the PEPS transfer matrix. This phenomenon became more prevalent at higher $D$ and only showed up for the Ising model. Finally, we always use the largest length scale in the boundary MPS. In the case of Heisenberg model it coincides with the spin-spin correlation length, but in other models it might prove more fruitful to use some smaller correlation length associated to a more relevant correlator; this can be achieved by using symmetric tensor networks or direct fitting of the relevant correlation function. 
\par Somewhat surprisingly, we have observed that the approximate nature of the environment was not 'abused' in the optimization to give energies below the ground state energy (or entropies/free energies above the physical one) as one might have expected. Whether this is universal or necessary for the validity of our scaling hypothesis remains unclear for now.

\par In conclusion, we have formulated a scaling hypothesis for PEPS and identified an induced correlation length $\xi(D,\chi)$ as the only relevant parameter when the PEPS is optimized using variational gradient methods at fixed bond dimensions $D$ and $\chi$. This allows for simulating quantum critical points in 2+1D with much greater accuracy and smaller cost. We conjecture that a similar scaling hypothesis holds in any dimension, therefore opening up the possibility of simulating quantum critical points in 3-D and higher.

\noindent\emph{Acknowledgments.}--- The authors would like to thank Fabien Alet and Andreas L\"{a}uchli for inspiring discussions. This work was supported by the Research Foundation Flanders (G0E1820N) and the ERC grant QUTE (647905). Part of the computations were carried out on the HPC resources of CALMIP supercomputing center under the allocation 2020-P1231. JH is supported by the TNSTRONG ANR-16-CE30-0025 and the TNTOP ANR-18-CE30-0026-01 grants awarded by the French Research Council.

%\bibliography{bibliography}

%merlin.mbs apsrev4-1.bst 2010-07-25 4.21a (PWD, AO, DPC) hacked
%Control: key (0)
%Control: author (0) dotless jnrlst
%Control: editor formatted (1) identically to author
%Control: production of article title (0) allowed
%Control: page (1) range
%Control: year (0) verbatim
%Control: production of eprint (0) enabled
%

\clearpage \newpage \clearpage
\appendix
\section*{Appendix}
\newcommand{\diagram}[2]{\quad\vcenter{\hbox{\includegraphics[scale=0.3,page=#2]{#1.pdf}}}\quad}

In the main body of the text, we have explained that it is crucial to differentiate the PEPS cost function evaluated at a fixed value of environment bond dimension $\chi$. In the context of two-dimensional quantum systems, differentiating the $\chi$-dependent cost function can be done through the technique of automatic differentiation, as shown in Refs.~\onlinecite{Liao2019} and \onlinecite{Hasik2020}. Here, we show that the same procedure is possible for PEPS approximations of transfer-matrix fixed points, which arise in the tensor-network representation of 3-D partition functions \cite{Vanderstraeten2018} or the imaginary-time evolution operator for 2-D quantum Hamiltonians \cite{Vanhecke2020}, under certain symmetry conditions on the involved tensors.
\par We start from a uniform projected entangled-pair operator (PEPO) for an infinite system, described by a 6-leg tensor $T$ as
\begin{equation}
	O(T) = \diagram{peps1}{7},
\end{equation}
which represents a layer of a three-dimensional tensor network. Evaluating this partition function amounts to finding the leading eigenvector $\ket{\Psi}$ of this transfer matrix, such that the free energy density is given by
\begin{equation}
	f = - \lim_{N\to\infty} \frac{1}{N} \log \left( \frac{ \bra{\Psi} O(T) \ket{\Psi}} {\braket{\Psi | \Psi }} \right),
\end{equation}
where $N$ is the diverging system size -- we will show that we can give meaning to the density in the thermodynamic limit directly, i.e. without taking the $N\to\infty$ limit. Assuming this PEPO is Hermitian, we will approximate the fixed point as a uniform PEPS, parametrized by a single complex-valued tensor $A$,
\begin{equation}
	\ket{\Psi(A)} = \diagram{peps1}{1},
\end{equation}
and solve the optimization problem
\begin{equation}
	A = \argmin_A f(A,\bar{A}),
\end{equation}
with
\begin{equation}
	f(A,\bar{A}) = - \frac{1}{N} \log \left( \frac{ \bra{\Psi(\bA)} T \ket{\Psi(A)}} {\braket{\Psi(\bA) | \Psi(A) }} \right)
\end{equation}
to find an optimal tensor $A$. Here, $\bA$ is the complex conjugate of the tensor $A$. In order to solve this optimization problem, we have to evaluate numerator and denominator in the above expression for a given PEPS, which boils down to contracting an infinite two-layer and three-layer tensor network. Here we use boundary MPS methods for the double-layer transfer matrix
\begin{equation}
	T_2 = \diagram{peps1}{4}, 
\end{equation}
with
\begin{equation}
	\diagram{peps1}{5} = \diagram{peps1}{6},
\end{equation}
and triple-layer transfer matrix
\begin{equation}
	T_3 = \diagram{peps1}{12},
\end{equation}
with
\begin{equation}
	\diagram{peps1}{11} = \diagram{peps1}{10},
\end{equation}
Here we assume that the tensor $T$ has the symmetry 
\begin{equation}
	\diagram{peps1}{8} = \diagram{peps1}{9}.
\end{equation}
with a given unitary matrix $X$, and that we have chosen the PEPS tensor with the symmetry constraint
\begin{equation}
	\diagram{peps1}{2} = \diagram{peps1}{3}
	\label{RT}
\end{equation}
with $X$ the same unitary matrix. These symmetry constraints can be generalised a bit, but the important thing is that the double-layer and triple-layer transfer matrix are both Hermitian operators. In the examples of the dimer and Ising model the $X$ matrix is a simple unit matrix.

\subsection*{Vumps equations}

We characterize the boundary MPS as the fixed point of the vumps (Variational Uniform Matrix Product State) algorithm; for more details on uniform MPS and the derivation of the fixed-point equations, see Ref.~\onlinecite{Vanderstraeten2019}. We first list the equations for a normalized MPS in the center gauge, described by four tensors $\{M_l,M_r,C,M_c\}$,
\begin{equation}
	\diagram{vumps}{11} = \diagram{vumps}{12} = \diagram{vumps}{13},
\end{equation}
with the conditions
\begin{equation}
	\diagram{vumps}{19} = \diagram{vumps}{20}, \qquad \diagram{vumps}{21} = \diagram{vumps}{22}
\end{equation}
and the normalization
\begin{equation}
	\diagram{vumps}{5}=1, \qquad \diagram{vumps}{14} = 1.
\end{equation}
We characterize the fixed point of a given hermitian MPO by the following fixed-point equations
\begin{equation}
	\diagram{vumps}{1} = \lambda \diagram{vumps}{2}
\end{equation}
and
\begin{equation}
	\diagram{vumps}{15} = \diagram{vumps}{16}.
\end{equation}
Note that the second equation for $C$ follows from the one for $M_c$. The left and right fixed points are determined by the following eigenvalue equations
\begin{align}
	\diagram{vumps}{6} = \lambda \diagram{vumps}{7}, \\
	\diagram{vumps}{8} = \lambda \diagram{vumps}{9},
\end{align}
with the normalization
\begin{equation}
	\diagram{vumps}{10} = 1.
\end{equation}
Importantly, through the hermiticity of the transfer matrix, we also have the fixed-point equations for $\bA_c$ and $\bar{C}$,
\begin{equation}
	\diagram{vumps}{3} = \lambda \diagram{vumps}{4},    
\end{equation}
and
\begin{equation}
	\diagram{vumps}{17} = \diagram{vumps}{18}.
\end{equation}

Given these fixed-point equations and normalizations, the transfer-matrix eigenvalue per site of the MPO is given by
\begin{equation}
	\lambda = \diagram{vumps}{23}
\end{equation}

\subsection*{Differentiating the fixed-point equations}

All tensors $\{M_l,M_r,M_c,C,\bar{M}_l,\bar{M}_r,\bar{M}_c,\bar{C},G_l,G_r\}$ can be thought of being functionally dependent on $O$ through the above fixed-point equations. If we change the transfer-matrix tensor as $O\to O+dO$, all these tensors have a first-order change as well. We can find these first-order tensors by differentiating the fixed-point equations. In the following we denote the first-order tensors with small letters.
\par We first differentiate the normalization conditions, arriving at the equalities
\begin{align}
	&\diagram{diff}{1} + \diagram{diff}{2} = 0, \\ 
	&\diagram{diff}{3} + \diagram{diff}{4} = 0.\label{Cnormdiff}
\end{align}
The normalization of $G_l$ and $G_r$ can be differentiated to give
\begin{multline}
	\diagram{diff}{5} + \diagram{diff}{6} + \diagram{diff}{7} \\ + \diagram{diff}{8} = 0  , 
\end{multline}
which, after using the eigenvalue equation for $C$ and $C^*$ and the equation \ref{Cnormdiff}, gives
\begin{equation}
	\diagram{diff}{5} + \diagram{diff}{6} = 0.
\end{equation}
Next, we find the first-order change in the transfer-matrix eigenvalue,
\begin{multline}
	d\lambda = \diagram{diff}{9} + \diagram{diff}{10} + \diagram{diff}{11} \\ + \diagram{diff}{12} + \diagram{diff}{13}.
\end{multline}
The four last terms can be simplified using the eigenvalue equations above and, therefore, cancel, so that we find
\begin{equation}
	d\lambda = \diagram{diff}{9}.
\end{equation}
This implies that the first-order change in the eigenvalue of the transfer matrix is given by the simple diagram where we differentiate the $O$ tensor.

\subsection*{Optimizing the free energy}

We now return to the objective function that we wish to optimize. Using the boundary MPS for double- and triple-layer transfer matrix, characterized by the vumps fixed-point equations, we approximate the free-energy density as
\begin{equation}
	f(A,\bA) = - \log \lambda(A,\bA),
\end{equation}
and
\begin{equation}
	\lambda = \diagram{opt1}{1} / \diagram{opt1}{2}.
\end{equation}
Here, the two-leg and three-leg boundary-MPS tensors denote different tensors. The function $\lambda(A,\bA)$ is a real function of complex parameters. The gradient of the objective function is given by
\begin{widetext}
	\begin{equation}
		g = - \frac{2}{\lambda(A,\bA)} \left( \diagram{opt1}{2} \right)^{-1} \left( \diagram{opt1}{3} - \lambda(A,\bA) \diagram{opt1}{4} \right) 
	\end{equation}
\end{widetext}
with
\begin{equation}
	e_2 = \partial_{\bA} E_2, \qquad e_3 = \partial_{\bA} E_3.
\end{equation}
Here, the differentials with respect to the environment tensors all vanish as we have showed above. This means that the gradient $g$ is the exact gradient for the objective function evaluated at a given value of the boundary MPS $\chi$. 

\end{document}